**Bennett, CC (2010). Clinical Productivity System – A Decision Support Model.** *International Journal of Productivity and Performance Management*. **60(3): 311-319.** http://www.emeraldinsight.com/journals.htm?articleid=1911824&show=abstract

# Clinical Productivity System – A Decision Support Model

Casey C. Bennett
Dept. of Informatics
Centerstone Research Institute
Nashville, TN

Contact Info:
Casey Bennett
44 Vantage Way
Suite 280
Nashville, TN 37208
U.S.A.
01-615-460-4111
Casey.Bennett@CenterstoneResearch.org

**Abstract**

*Purpose*: This goal of this study was to evaluate the effects of a data-driven clinical productivity system that leverages Electronic Health Record (EHR) data to provide productivity decision support functionality in a real-world clinical setting. The system was implemented for a large behavioral health care provider seeing over 75,000 distinct clients a year.

*Design/methodology/approach*: The key metric in this system is a "VPU", which simultaneously optimizes multiple aspects of clinical care. The resulting mathematical value of clinical productivity was hypothesized to tightly link the organization's performance to its expectations and, through transparency and decision support tools at the clinician level, affect significant changes in productivity, quality, and consistency relative to traditional models of clinical productivity.

*Findings*: In only 3 months, every single variable integrated into the VPU system showed significant improvement, including a 30% rise in revenue, 10% rise in clinical percentage, a 25% rise in treatment plan completion, a 20% rise in case rate eligibility, along with similar improvements in compliance/audit issues, outcomes collection, access, etc.

*Practical implications*: A data-driven clinical productivity system employing decision support functionality is effective because of the impact on clinician behavior relative to traditional clinical productivity systems. Critically, the model is also extensible to integration with outcomes-based productivity.

*Originality/Value:* EHR's are only a first step - the problem is turning that data into useful information. Technology can leverage the data in order to produce actionable information that can inform clinical practice and decision-making. Without additional technology, EHR's are essentially just copies of paper-based records stored in electronic form.

**I. Introduction**

Recent years have the seen the proliferation of electronic health records (EHR's) across the mental healthcare field. The problem is turning that data into useful information. The actual collection of data in an electronic health record (EHR) is only the first step – indeed, we must leverage that data through technology in order to provide useable, actionable information. Even the popular media is picking up on this fact ("Little Benefit Seen, So Far, in Electronic Patient Records" *New York Times*, 11/15/2009). Without additional technology above and beyond simply collecting and displaying information, EHR's are essentially just copies of paper-based records that happen to be stored in electronic form. The expected gain in terms of clinical outcomes, quality, and efficiency is limited.

Here, we describe one example of this approach – a clinical productivity system incorporating decision support methodology, based on mathematical models of productivity and efficiency and utilizing a feedback mechanism of decision support (Wan, 2006; March and Hevnar, 2007; Wright and Sittig, 2008). The productivity metric is completely data driven, rather than based on an arbitrary weighting system. This data driven approach allows the productivity system to drive productivity as well as quality, with the potential to be adapted to drive clinical outcomes as well. The result of this approach is a tightly bound mathematical relationship that directly links actual production to the budgeted expectation, as suggested in the theoretical literature (Purbey et al., 2007). In order to evaluate the system's effectiveness, we compare it to traditional models of clinical productivity in a real-world setting, evaluating the effects pre versus post of a system-wide implementation.

Historically, productivity systems and related research have attempted to assign this value using such key concepts as "weighted productivity", billable hours, and RVU's (relative value units). By far, the most common methods to measure clinical productivity have used a "weighted productivity" metric based on weighted billable hours (Abouleish, 2008). This method assigns some approximation of the "value" of a clinical service. For instance, an individual therapy session might receive a '1' while a group therapy gets a '0.5' based on some perceived 2:1 value ratio between the two. The problem is that, in reality, those weightings are in essence arbitrary and lack any mathematical foundation to accurately "value" a clinical service. Thus the ratios, which are intended to influence clinical choices, are misaligned. One proposed improvement to these "weighted productivity" systems was the concept of RVU's (Glass and Anderson, 2002; Willis et al., 2004). RVU's are still weighted measures, but are based on Medicare recommended rates. Compared to the predecessor systems based on arbitrary valuations, RVU's represent a positive movement towards constructing a mathematical framework around clinical productivity. However, RVU's still face limitations in that they impact only quantity rather than quality of care (Willis et al., 2004).

## II. Methods
### A. Setting

Centerstone is the largest community-based behavioral healthcare provider in the United States, seeing over 75,000 distinct clients a year in both Tennessee and Indiana. Centerstone Research Institute is an arm of that organization devoted to integrating evidence and practice, conducting clinical research, developing clinical decision support tools, and building new healthcare informatics technologies. Centerstone has a fully functional EHR that maintains virtually all relevant patient records.

### B. System Description

The productivity system is built around the concept of a VPU (value per unit). These values are based initially on revenue, or in situations where that is not always clear, estimated revenue (typically this revenue is actually averaged across payers for each service, for compliance reasons). It is important to note that the productivity system links directly to the billing system in order to ensure that valid claims are created for each service and payer requirements met. Other variables, such as quality of care measures, treatment plan completion, and eligibility issues (e.g. appropriate staff licensure), are factored in as modifiers to create the final productivity metric. This final metric "values" each individual service in the system based on the data. In order to facilitate clinician efficiency, this information is fed back on a daily basis, using the "feedback" model of decision support (Wan, 2006; March and Hevnar, 2007; Wright and Sittig, 2008). The system is based completely on open-source technology, and the design is considered non-proprietary.

Two other key variables are Clinical Full Time Equivalent (Clinical FTE), which is the portion of staff's total FTE expected to be clinical and allows adjustments for staff with supervisory/management responsibilities, and Clinical Percentage, which is the percentage of time a clinician is expected to spend providing direct, face-to-face, billable client services and allows adjustments for the administrative (e.g. paperwork) and other non-revenue generating duties associated with provision of actual clinical services by clinicians. For most staff, clinical percentage is set at 62.5%, which equates to 100-105 billable hours a month, or roughly 5 hours a day.

The mathematical equations were structured so that all staff in effect have the same target: 100 VPU's per month per 1.0 clinical FTE. This simplifies management, and allows for staff doing completely different jobs to be directly compared based on their productivity percentage, which is simply actual credit divided by the expected target.

The VPU credit is in essence a formula. The "moveable levers" in this formula are the <u>expected clinical percentage</u> and <u>expected monthly revenue</u> for each clinician. The expectations (and thus the budget) change, not the VPU's themselves. The VPU's are calculated based off actual performance versus expected performance in order to ensure a tightly bound relationship between the two. The principle formula can be represented as follows:

$R_e$ = Expected Monthly Revenue for Staff = $9,000

$CP$ = Clinical Percent = 62.5%

$H_t$ = Total Work Hours per Month = 160 (for 1.0 FTE)

$R_a$ = Actual Revenue for Service = $100

Then the calculation is:

$VPU_{base} = R_a / (R_e / (H_t * CP))$

Ex. - $VPU_{base} = \$100 / (\$9000 / (160 * .625)) = .9$

Where the numerator = actual production, and the divisor = budgeted expectation. Finally:

$VPU_{final} = VPU_{base} *$ Modifiers

The method of applying specific modifiers varies by metric, depending on the intent relative to clinical behavior.

Given the study was interested in the real-world application of the theoretical constructs of clinical productivity to an actual productivity system, the principle form of comparison is the

Clinical Productivitypre versus post analysis of the metrics incorporated into the productivity system against the previous "weighted" system, including measures of productivity, quality, and efficiency. Although some suggest that real-world studies lack the rigor and control of systematic studies (Balas and Boren, 2007), controlled, systematic studies lack the variability of real-world settings that typically undermine the application of theoretical frameworks (Kaplan, 2001; Rahimi and Vimarlund, 2007).

## III. Results

The productivity system described here simultaneously optimizes health care organizational function around a number of domains. The impact was significant. Note that the time ranges and scales on the subsequent graphs and tables vary due to availability of the underlying data and the different launch dates for the system's various components. The data period presented was limited by other organizational changes (e.g. contractual changes at either end) and because historical analysis of that range indicated no major seasonal affects during the period in question. In short, implementation studies are difficult in real-world settings, but the time period in question was the best available.

The most immediate impact of the VPU clinical productivity system was on revenue and how staff scheduled their time (Figure 1).

**Figure 1: Monthly Revenue**

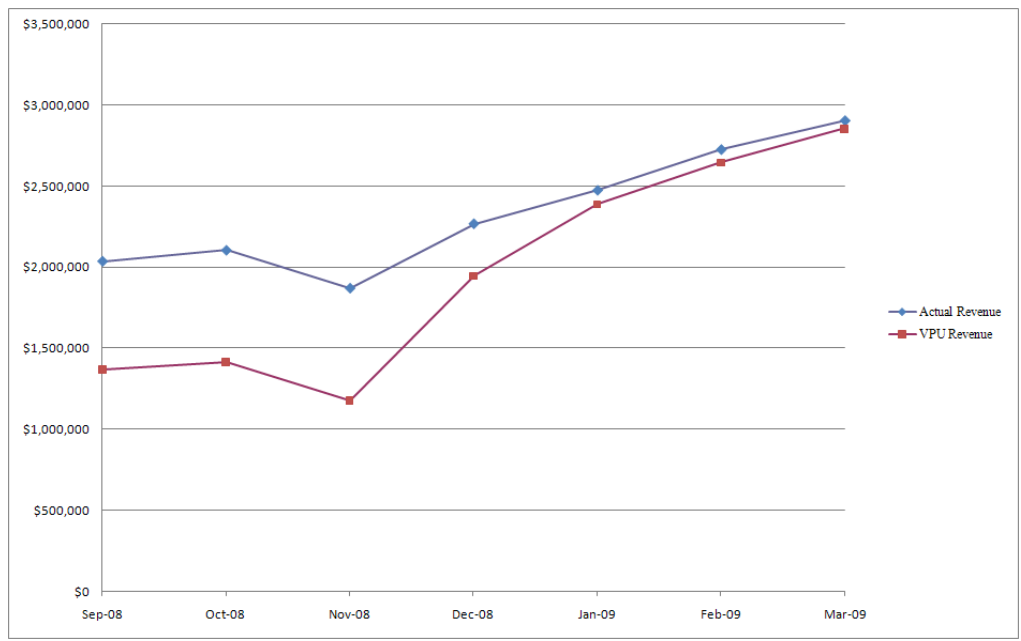

Using December 2008 as a baseline till March 2009, or just 3 months, there was a 28% increase in Actual Revenue. Using a paired *t*-test, the March average (*M*=$7,528 per staff) was significantly greater than the December average (*M*=$5,409 per staff), *t*(344)=-14.071, *p*<.001. Total staff time only rose 2.1% during that same period and actually fell 7.9% between December and February (during which there was still a 21.7% rise in revenue). Historically, revenue has remained fairly stable between September and May each year (data not shown). Also of note from Figure 1 was the decrease in variation between the Actual Revenue and the VPU Revenue, where the initial variation fell from nearly 30% to 1-3%. This disparity was due to a misalignment of disparate targets and ancillary staff responsibilities necessary to actually get paid and/or avoid compliance/audit issues.

The VPU productivity system also had a positive impact on a number of ancillary components of the system as they related to client care, collection of outcomes, and compliance/audit issues. For example, treatment plan completion rates increased approximately 25% (data not shown). Using paired *t*-test, the April average (*M*=94%) was significantly greater than the September average (*M*=68.8%), *t*(306)=-14.929, *p*<.001.

Additionally, the productivity system had a major impact on case management services (Figure 2). Note that different programs adopted use in different months – the first adopters being Child and Youth (C&Y) case management in October 2008. Adult areas began actively using during November and December of 2008. School-based case managers were the last adopters. The pattern of adoption can clearly be seen in the graph. The average case management eligibility in September was 79.4%, with a standard deviation of 7%. By March 2009, the average was 96.3% with a standard deviation of 1%. Not only has the average significantly increased (paired *t*-test, *t*(157)=-11.325, *p*<.001), but unnecessary variation was driven out of the system. Figure 3 shows the absolute increase in eligible cases during that time period relative to the overall client count in case management (eligibility here was determined based on Medicaid [Tenncare] which represented 90% of clients in case management). In September 2008 total caseload of 4,519. By March 2009, total <u>eligible clients</u> exceeded baseline total caseload by 10.6%. In other words, Centerstone was getting paid for over 10% more case management clients than were even on caseloads to begin with.

**Figure 2: Case Management Eligible Cases**

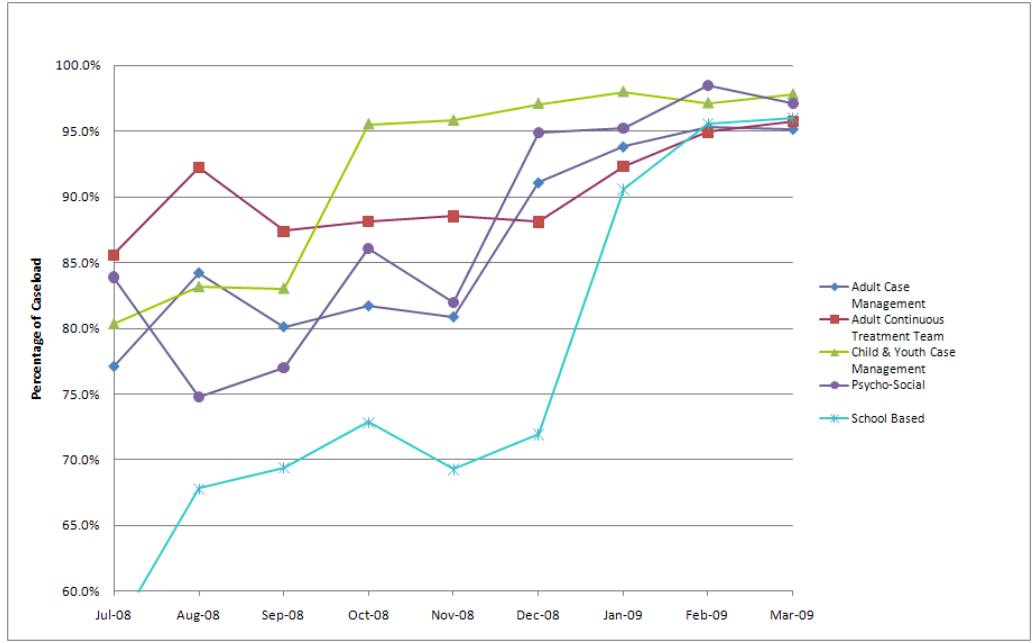

Figure 3: Case Management Client Count

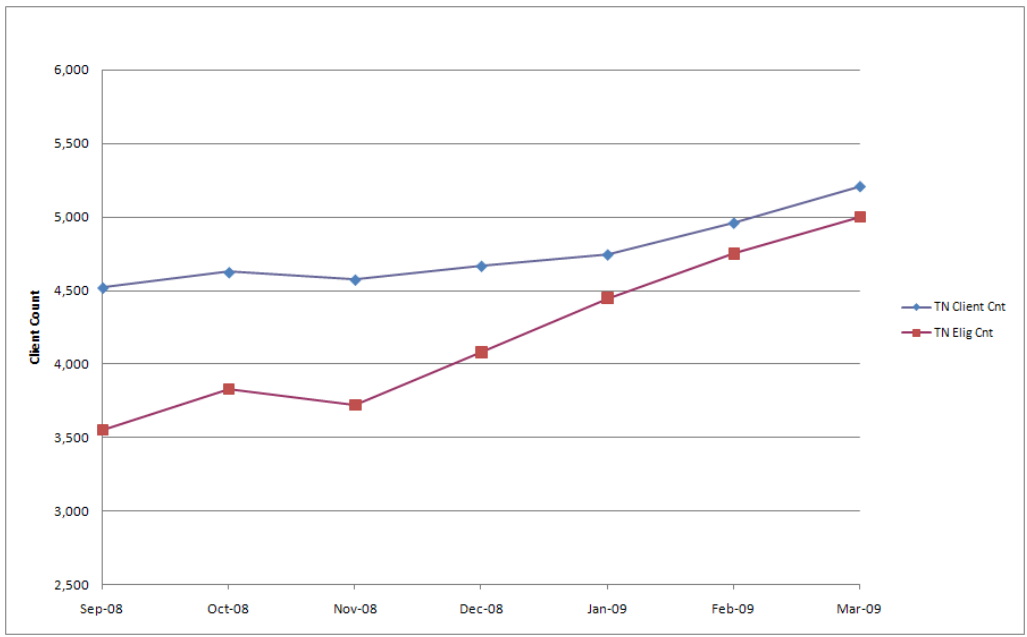

Finally, there is some assumption that this kind of productivity system might decrease access for certain groups based on payment ability. In fact, the VPU productivity system has coincided with a 23% decrease in access time across the board (defined as time from scheduling of intake to actual intake, data not shown). Access time for consumers with Tenncare (Medicaid), Safety Net, and Self Pay are all down significantly – in fact they are at historical lows, as is access time overall. Overall, increased efficiency in service provision appears to coincide with decreases in access intervals.

**IV. Discussion**

The VPU clinical productivity system has been very successful to-date in affecting change within the organization. This change can be seen in the various improvements in metrics related to revenue, billable hours, clinical percentage, compliance/audit variables, caseload expectations, chart completion, authorizations, outcomes collection, access times, etc. These changes are the result of leveraging the data in the organization's EHR to inform clinical decision-making. The results also suggest that a data-driven approach to clinical productivity is potentially more effective than a historical "weighted" productivity methodology.

There are significant limitations in the current work, primarily due to the real-world implementation of the system. Given that the system grew out of the need to address pressing revenue and productivity concerns in live clinical practice, a rigorous study design was impractical. Additional work remains to evaluate the effects of such an approach to clinical productivity in controlled, systematic settings. The model also currently lacks any quantitative or qualitative evaluation of specific behavioral changes of affected clinicians, which is a critical barrier to real-world implementation (Francke et al., 2008).

A data-driven, mathematical approach to clinical productivity holds potential utility for healthcare providers. First, the mathematical approach opens the door to the inclusion of outcome measures directly into productivity metrics, rather than utilizing two separate metrics. This is important because better outcomes may sometimes be in conflict with higher revenue and/or billable time. Integrating these into one measure can capture the sometimes synergistic, sometimes conflicting, nature of producing them, allowing for simultaneous optimization of both. Second, this inclusion of outcomes can be advantageous in pay-for-performance environments by driving both efficiency and quality. It is likely that the lack of such clinical productivity measures have impeded prior attempts to sustain pay-for-performance models (Peterson et al., 2006; Rosenthal and Frank, 2006; Rosenthal, 2008; McDonald and Roland, 2009). A third advantage is that the productivity system links directly into the billing system, and thus captures the conflicting payment methodologies of different payers, sometimes even for the same exact service. This is critical in a hybrid fee-for-service and case rate environment, as otherwise clinicians must rely on ad-hoc information in order to meet payer rules and restrictions and often provide unnecessary and/or unbillable services.  Finally, this approach reduces overall variability in client care.  Incorporating various quality measures, such as the requirement of completed treatment plans, directly into the productivity system via data calculations enhances compliance.

A long-term goal is to mathematically incorporate client outcomes directly into a clinical productivity metric and optimize them simultaneously with all other variables.  This would allow a clinical productivity system to adapt to any payment methodology, including a pay-for-performance methodology or an outcomes-incentive model.  Outcomes would be applied as a variable in the equation by converting them into a cost-based value reverse-engineered from the data itself, calculating "cost per unit change" (CPUC) from the total population using a standardized rates and/or deriving CPUC from population metrics (e.g. hospitalization rates, etc.).  After that, the value can be applied as scaling factor.  Another possibility would be to calculate the delta, Δ, per client per month (change from baseline to end point).  As an example:

$H_e$ = Expected Hourly Earnings = $100

CPUC = Cost Per Unit Change = $500

$C_h$ = Hours of Client's Services that Month = 4.5

$O_0$ = Outcome at Baseline = 2.5

$O_1$ = Outcome at End Point = 3.5

Then the slicer (S) is:

$S = (CPUC * (O_1 - O_0)) / (H_e * C_h) = (500 * (3.5 - 2.5)) / (4.5*100) = 1.11$

So for one particular service that month, if:

$VPU_{base}$ = Base VPU = .9

Then:

$VPU_{base} * S = .9 * 1.11 = .99 = VPU_{final}$

Where $O_1 - O_0$ equates to $\Delta$. This equation is simplified for explanatory purposes. There may be challenges to incorporating outcomes into these equations depending on the funding environment, where better outcomes may not be financially incentivized.

        The future is moving towards the integration of outcomes and clinical quality data with productivity across the healthcare spectrum, coinciding with pressures to hold down costs while improving the quality of care. The adoption of EHR's is indeed only the first step. Technology must be developed to leverage the data existing within those EHR's, producing actionable information that can inform clinical practice and make predictions about future events (Bennett and Doub, 2010). These technologies can turn EHR's into the transformative decision support tools they were envisioned as in the beginning.

**Acknowledgements**

This project was funded by grants through the Ayers Foundation and the Joe C. Davis Foundation.  The funders had no role in design, implementation, or analysis of this research.  The author would also like to recognize various Centerstone staff for their contributions to this effort: Dr. Tom Doub, Vice President Ben Middleton, among others.  Special thanks to Christina Van Regenmorter at Centerstone Research Institute for draft review assistance and manuscript preparation.  This paper was presented at Mental Health Services Research Conference 2009, Washington, DC.  The author has no direct financial interest in the productivity system, nor is the system intended for purchase.  Most of the software and computing environment is open-source (Pentaho, Jasper, Postgres, etc.) by intention, so that the functionality of the system does not depend on any propietary technology.

**References**

Abouleish, A.E. (2008), "Productivity-based compensations versus incentive plans", *Anesthesia & Analgesia*, Vol.107, No.6, pp.1765-1767.

Balas, E. and Boren, S. (2007), "Clinical Trials of Information Interventions", Berner, E. (ed), *Clinical Decision Support Systems: Theory and Practice (2$^{nd}$ ed),* Springer, New York, pp.140-155.

Bennett, C.C. and Doub, T.W. (2010), "Data Mining and Electronic Health Records: Selecting Optimal Clinical Treatments in Practice", in *Proceedings of the 6th International Conference on Data Mining, Las Vegas, USA,* CSREA Press, pp.313-318.

Francke, A.L., Smit, M.C., de Veer, A.J. and Mistiaen, P. (2008), "Factors influencing the implementation of clinical guidelines for health care professionals: a systematic meta-review", *BMC Medical Informatics and Decision Making*, Vol.8, No.1, pp.38-49.

Glass, K. and Anderson, J. (2002), "Relative value units: from A to Z", *Journal of Medical Practice Management,* Vol.7, No.5, pp.225-228.

Kaplan, B. (2001), "Evaluating informatics applications--clinical decision support systems literature review", *International Journal of Medical Informatics*, Vol.64, No.1, pp.15-37.

March, S.T. and Hevner, A.R. (2007), "Integrated decision support systems: A data warehousing perspective", *Decision Support Systems*, Vol.43, No.3, pp.1031–1043.

McDonald, R. and Roland, M. (2009), "Pay for Performance in Primary Care in England and California: Comparison of Unintended Consequences", *Annals of Family Medicine*, Vol.7, No.2, pp.121-127.

Petersen, L.A., Woodard, L.D., Urech, T., Daw, C. and Sookanan, S. (2006), "Does Pay-for-Performance Improve the Quality of Health Care?" *Annals of Internal Medicine*, Vol.145, No.4, pp.265-272.

Purbey, S., Mukherjee, K. and Bhar, C. (2007), "Performance measurement system for healthcare processes", *International Journal of Productivity and Performance Management,* Vol.56, No.3, pp.241-251.

Rahimi, B. and Vimarlund, V. (2007), "Methods to Evaluate Health information Systems in Healthcare Settings: A Literature Review", *Journal of Medical Systems*, Vol.31, No.5, pp.397-432.

Rosenthal, M.B. and Frank, R.G. (2006), "What Is the Empirical Basis for Paying for Quality in Health Care?" Medical *Care Research Review*, Vol.63, No.2, pp.135-157.

Rosenthal, M.B. (2008), "Beyond Pay for Performance -- Emerging Models of Provider-Payment Reform", *New England Journal of Medicine*, Vol.359, No.12, pp.1197-1200.

Wan, T. (2006), "Healthcare Informatics Research: From Data to Evidence-Based Management", *Journal of Medical Systems*, Vol.30, No.1, pp.3-7.

Willis, D., Kelton, G., Saywell, R. and Kiovsky, R. (2004) "An incentive compensation system that rewards individual and corporate productivity", *Family Medicine*, Vol.36, No.4, pp.270-278.

Wright, A. and Sittig, D.F. (2008), "A four-phase model of the evolution of clinical decision support architectures", *International Journal of Medical Informatics*. Vol.77, No.10, pp.641–649.